\documentclass[aps,amsfonts,amsmath,amssymb,nofootinbib,twocolumn,superscriptaddress]{revtex4}

\usepackage{graphics}
\usepackage{hyperref}
\usepackage{epsfig}
\usepackage{color}

\usepackage{multirow}
\usepackage{bm}

\def\t#1{\textrm{#1}}
\def\ket#1{|#1\rangle }
\def\bra#1{\langle #1 |}

\def\n{\nonumber \\ }

\begin{document}

\title{Giant anisotropic nonlinear optical response \\ in transition metal monopnictide Weyl semimetals}



\author{Liang Wu}
\email{liangwu@berkeley.edu}
\affiliation{Department of Physics, University of California, Berkeley, California 94720, USA}
\affiliation{Materials Science Division, Lawrence Berkeley National Laboratory, Berkeley, California 94720, USA}
\author{S. Patankar}
\affiliation{Department of Physics, University of California, Berkeley, California 94720, USA}
\affiliation{Materials Science Division, Lawrence Berkeley National Laboratory, Berkeley, California 94720, USA}
\author{T. Morimoto}
\affiliation{Department of Physics, University of California, Berkeley, California 94720, USA}
\author{N. L. Nair}
\affiliation{Department of Physics, University of California, Berkeley, California 94720, USA}
\author{E. Thewalt}
\affiliation{Department of Physics, University of California, Berkeley, California 94720, USA}
\affiliation{Materials Science Division, Lawrence Berkeley National Laboratory, Berkeley, California 94720, USA}
\author{A. Little}
\affiliation{Department of Physics, University of California, Berkeley, California 94720, USA}
\affiliation{Materials Science Division, Lawrence Berkeley National Laboratory, Berkeley, California 94720, USA}
\author{J. G. Analytis}
\affiliation{Department of Physics, University of California, Berkeley, California 94720, USA}
\affiliation{Materials Science Division, Lawrence Berkeley National Laboratory, Berkeley, California 94720, USA}
\author{J. E. Moore}
\affiliation{Department of Physics, University of California, Berkeley, California 94720, USA}
\affiliation{Materials Science Division, Lawrence Berkeley National Laboratory, Berkeley, California 94720, USA}
\author{J. Orenstein}
\email{jworenstein@lbl.gov}
\affiliation{Department of Physics, University of California, Berkeley, California 94720, USA}
\affiliation{Materials Science Division, Lawrence Berkeley National Laboratory, Berkeley, California 94720, USA}

 \date{\today}


\maketitle

\textbf{Although Weyl fermions have proven elusive in high-energy physics, their existence as emergent quasiparticles has been predicted in certain crystalline solids in which either inversion or time-reversal symmetry is broken\cite{WanPRB2011,BurkovPRL2011, WengPRX2015,HuangNatComm2015}. Recently they have been observed in transition metal monopnictides (TMMPs) such as TaAs, a class of noncentrosymmetric materials that heretofore received only limited attention \cite{XuScience2015, LvPRX2015, YangNatPhys2015}. The question that arises now is whether these materials will exhibit novel, enhanced, or technologically applicable electronic properties. The TMMPs are polar metals, a rare subset of inversion-breaking crystals that would allow spontaneous polarization, were it not screened by conduction electrons \cite{anderson1965symmetry,shi2013ferroelectric,kim2016polar}. Despite the absence of spontaneous polarization, polar metals can exhibit other signatures of inversion-symmetry breaking, most notably second-order nonlinear optical polarizability, $\chi^{(2)}$, leading to phenomena such as optical rectification and second-harmonic generation (SHG). Here we report measurements of SHG that reveal a giant, anisotropic $\chi^{(2)}$ in the TMMPs TaAs, TaP, and NbAs. With the fundamental and second harmonic fields oriented parallel to the polar axis, the value of $\chi^{(2)}$ is larger by almost one order of magnitude than its value in the archetypal electro-optic materials GaAs \cite{bergfeld2003second} and ZnTe \cite{wagner1998dispersion}, and in fact larger than reported in any crystal to date.}

The last decade has witnessed an explosion of research investigating the role of bandstructure topology, as characterized for example by the Berry curvature in momentum space, in the electronic response functions of crystalline solids \cite{NiuRMP2010}.  While the best established example is the intrinsic anomalous Hall effect in time-reversal breaking systems \cite{NagaosaRMP2010}, several nonlocal \cite{Orenstein2013,Zhong2015} and nonlinear effects related to Berry curvature generally \cite{MoorePRL2010,SodemannPRL2015}and in WSM's specifically \cite{Ishizuka2016,Chan2016} have been predicted in crystals that break inversion symmetry. Of these, the most relevant to this work is a theoretical formulation \cite{Morimoto2015} of SHG in terms of the "shift vector," which is a quantity related to the difference in Berry connection between two bands that participate in an optical transition.

\begin{figure*}[htp]
\includegraphics[width=\textwidth]{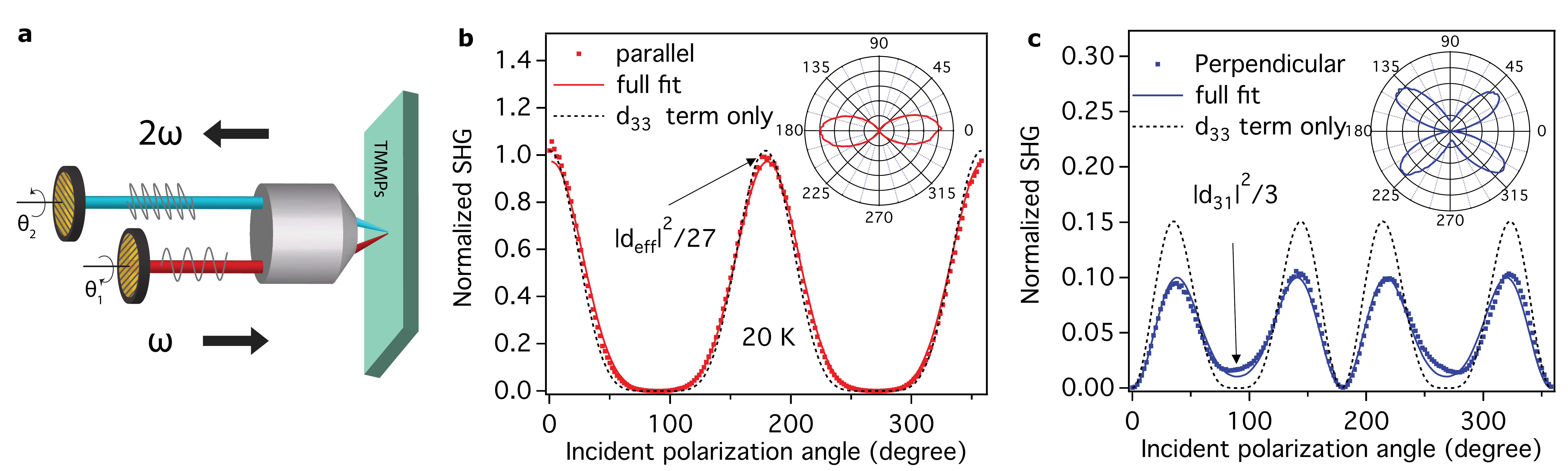}

\caption{  \textsf{\textbf{Second harmonic generation vs. angle as TaAs is effectively rotated about the axis perpendicular to the (112) surface.}} (\textsf{\textbf{a}}), Schematic of SHG experimental setup. To stimulate second harmonic light, pulses of 800 nm wavelength ($\omega$ frequency) were focused at near normal incidence to a 10 micron diameter spot on the sample.  Polarizers and waveplates mounted on rotating stages allowed for continuous and independent control of the polarization of the generating and  second-harmonic (2$\omega$ frequency) light that reached the detector.  $\theta_1$ and $\theta_2$ are the angle of the polarization plane after the generator and the analyzer respectively with respect to the [1,1,-1] crystal axis. (\textsf{\textbf{b}}), (\textsf{\textbf{c}}) SHG intensity as a function of angle of incident polarization at 20 K. In (\textsf{\textbf{b}}) and (\textsf{\textbf{c}}) the analyzer is parallel and perpendicular to the generator, respectively. In both plots the amplitude is normalized by the peak value in (\textsf{\textbf{b}}). The solid lines are fits that include all the allowed tensor elements in $4mm$ symmetry, whereas the dashed lines include only $d_{33}$. The insets are polar plots of measured SHG intensity vs. incident polarization angle. }
 \label{Fig1}
\end{figure*}

\begin{figure*}[htp]
\includegraphics[width=\textwidth]{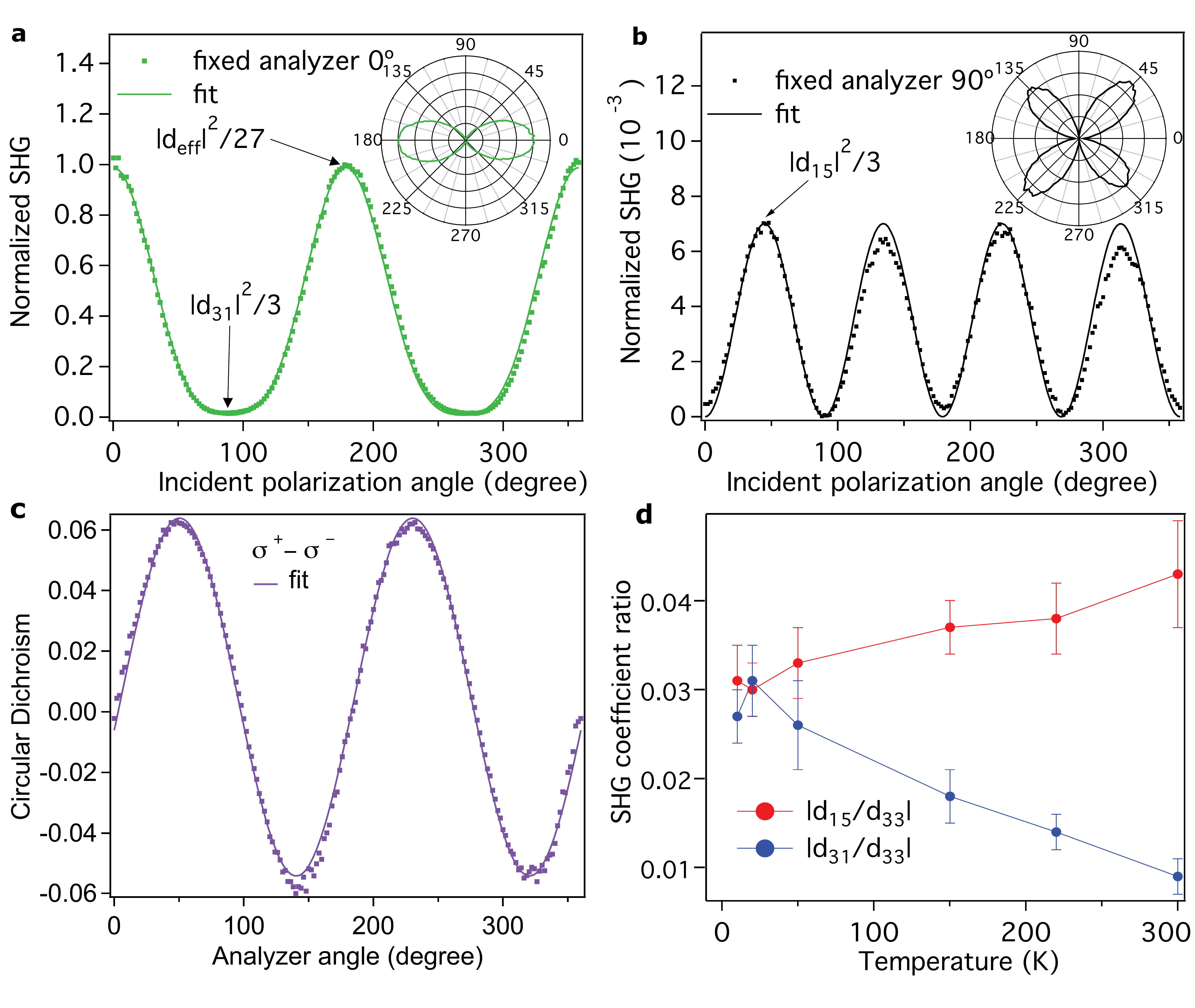}
\caption{  \textsf{\textbf{Second harmonic intensity with fixed analyzers, circular dichroism and temperature dependence on a TaAs (112) sample.}}  SHG as function of generator polarization with analyzers fixed along (\textsf{\textbf{a}}) 0 $^{\circ}$ and (\textsf{\textbf{b}})  90 $^{\circ}$, which are [1,1,-1] and [1,-1,0] crystal axes, respectively. Solid lines are fits considering all three non-zero tensor elements.  The insets are polar plots of measured SHG intensity. (\textsf{\textbf{c}}) Circular dichroism (SHG intensity difference between right- and left-hand circularly polarized generation light) normalized by the peak value in Fig. 1(\textsf{\textbf{b}}) \textit{vs}. polarization angle of the analyzer at 300 K. The solid line is a fit. (\textsf{\textbf{d}}) Temperature dependence of $|d_{15}/d_{33}|$ and $|d_{31}/d_{33}|$ of a TaAs (112) sample. The error bar is determined by setting $d_{15}$ and $d_{31}$ in and out of phase to $d_{33}$ in the fits.}
 \label{Fig2}
\end{figure*}

Fig. \ref{Fig1}a and caption provide a schematic and description of the optical set-up for measurement of SHG in TMMP crystals. Figs. \ref{Fig1}b,c show results from a (112) surface of TaAs. The SH intensity from this surface is very strong, allowing for polarization rotation scans with signal-to-noise ratio above 10$^6$. In contrast, SHG from a TaAs (001) surface is barely detectable (at least six orders of magnitude lower than (112) surface). Below, we describe the use of the set-up shown in Fig. \ref{Fig1}a to characterize the second-order optical susceptibility tensor, $\chi_{ijk}$, defined by the relation, $P_i(2\omega)=\epsilon_0 \chi_{ijk}(2\omega)E_j(\omega)E_k(\omega)$.

As a first step, we determined the orientation of the high-symmetry axes in the (112) surface, which are the [1,-1,0] and [1,1,-1] directions. To do so, we simultaneously rotated the linear polarization of the generating light (the generator) and the polarizer placed before the detector (the analyzer), with their relative angle set at either 0$^\circ$ or 90$^\circ$.  Rotating the generator and analyzer together produces scans, shown in Figs. \ref{Fig1}b, c, which are equivalent to rotation of the sample about the surface normal. The angles at which we observe the peak and the null in Fig. \ref{Fig1}b and the null in Fig. \ref{Fig1}c allow us to identify the principle axes in the (112) plane of the surface.

Having determined the high symmetry directions, we characterize $\chi_{ijk}$ by performing three of types of scans, the results of which are shown in Fig. \ref{Fig2}. In scans shown in Figs. \ref{Fig2}a, b, we oriented the analyzer along one of the two high symmetry directions and rotated the plane of linear polarization of the generator through 360$^\circ$. Fig. \ref{Fig2}c shows the circular dichroism of the SHG response, that is, the difference in SH generated by left and right circularly polarized light. For all three scans the SHG intensity as a function of angle is consistent with the second-order optical susceptibility tensor expected for the $4mm$ point group of TaAs, as indicated by the high accuracy of the fits in Fig. \ref{Fig1}b, c and Fig. \ref{Fig2}a-c.

In the $4mm$ structure $xz$ and $yz$ are mirror planes but reflection through the $xy$ plane is not a symmetry; therefore TaAs is an acentric crystal with an unique polar ($z$) axis. In crystals with $4mm$ symmetry there are three independent nonvanishing elements of $\chi_{ijk}$: $\chi_{zzz}$, $\chi_{zxx}=\chi_{zyy}$, and $\chi_{xzx}=\chi_{yzy}=\chi_{xxz}=\chi_{yyz}$. Note that each has at least one $z$ component, implying null electric dipole SHG when all fields are in the $xy$ plane. This is consistent with observation of nearly zero SHG for light incident on the (001) plane. Below, we follow the convention of using the $3\times 6$ second-rank tenor $d_{ij}$, rather than $\chi_{ijk}$, to express the SHG response, where the relation between the two tensors for TaAs is: $\chi_{zzz}=2d_{33}$, $\chi_{zxx}=2d_{31}$, and $\chi_{xzx}=2d_{15}$\cite{Boyd} (See Methods and Supplementary Information section A).

Starting with the symmetry constrained $d$ tensor, we derive expressions, specific to the (112) surface, for the angular scans with fixed analyzer shown in Fig. 2 a,b (Methods and Supplementary Information section A). We obtain $|d_{\textrm{eff}}\cos^2\theta_1+3d_{31}\sin^2\theta_1|^2/27 $ and $|d_{15}|^2\sin(2\theta_1)^2/3$ for analyzer parallel to [1,1,-1] and [1,-1,0], respectively, where $d_{\textrm{eff}}\equiv d_{33}+2d_{31}+4d_{15}$. Fits to these expressions yield two ratios: $|d_{\textrm{eff}}/d_{15}|$ and $|d_{\textrm{eff}}/d_{31}|$. Although we do not determine $|d_{33}/d_{15}|$ and $|d_{33}/d_{31}|$ directly, it is clear from the extreme anisotropy of the angular scans that $d_{33}$, which gives the SHG response when both generator and analyzer are parallel to the polar axis, is much larger than the other two components. We can place bounds on $|d_{33}/d_{15}|$ and $|d_{33}/d_{31}|$ by setting $d_{15}$ and $d_{31}$ in and out of phase with $d_{33}$. We note that the observation of circular dichroism in SHG, shown in Fig. \ref{Fig2}c, indicates that relative phase between $d_{15}$ and $d_{33}$ is neither $0^{\circ}$ or $180^{\circ}$, but rather closer to $30^{\circ}$ (Supplementary Information section A).

The results of this analysis are plotted in Fig. \ref{Fig2}d, where it is shown that $|d_{33}/d_{15}|$ falls in the range $\sim 25-33$ for all temperatures, and $|d_{33}/d_{31}|$ increases from $\sim 30$ to $\sim100$ with increasing temperature. Perhaps because of its polar metal nature, the anisotropy of the second-order susceptibility in TaAs is exceptionally large compared with what has been observed previously in crystals with the same set of nonzero $d_{ij}$. For example, $\alpha$-ZnS, CdS, and KNiO$_3$ have $|d_{31}|\cong|d_{15}|\cong d_{33}/2$ \cite{shoji1997absolute}, while in BaTiO$_3$ the relative sizes are reversed, with $|d_{31}|\cong |d_{15}|\approx 2 |d_{33}|$ \cite{miller1964optical}.

Even more striking than the extreme anisotropy of $\chi_{ijk}$ is the absolute size of the SHG response in TaAs.  The search for materials with large second-harmonic optical susceptibility has been of continual interest since the early years of nonlinear optics \cite{bloembergen1962light1}. To determine the absolute magnitude of the $d$ coefficients in TaAs we used GaAs and ZnTe as benchmark materials. Both crystals have large and well characterized second-order optical response functions \cite{bergfeld2003second, wagner1998dispersion}, with GaAs regarded as having among the largest $\chi^{(2)}_{ijk}$ of any known material. GaAs and ZnTe are also ideal as benchmarks because their response tensors have only one nonvanishing coefficient, $d_{14}\equiv 2\chi_{xyz}$.

\begin{figure*}[htp]

\includegraphics[width=\textwidth]{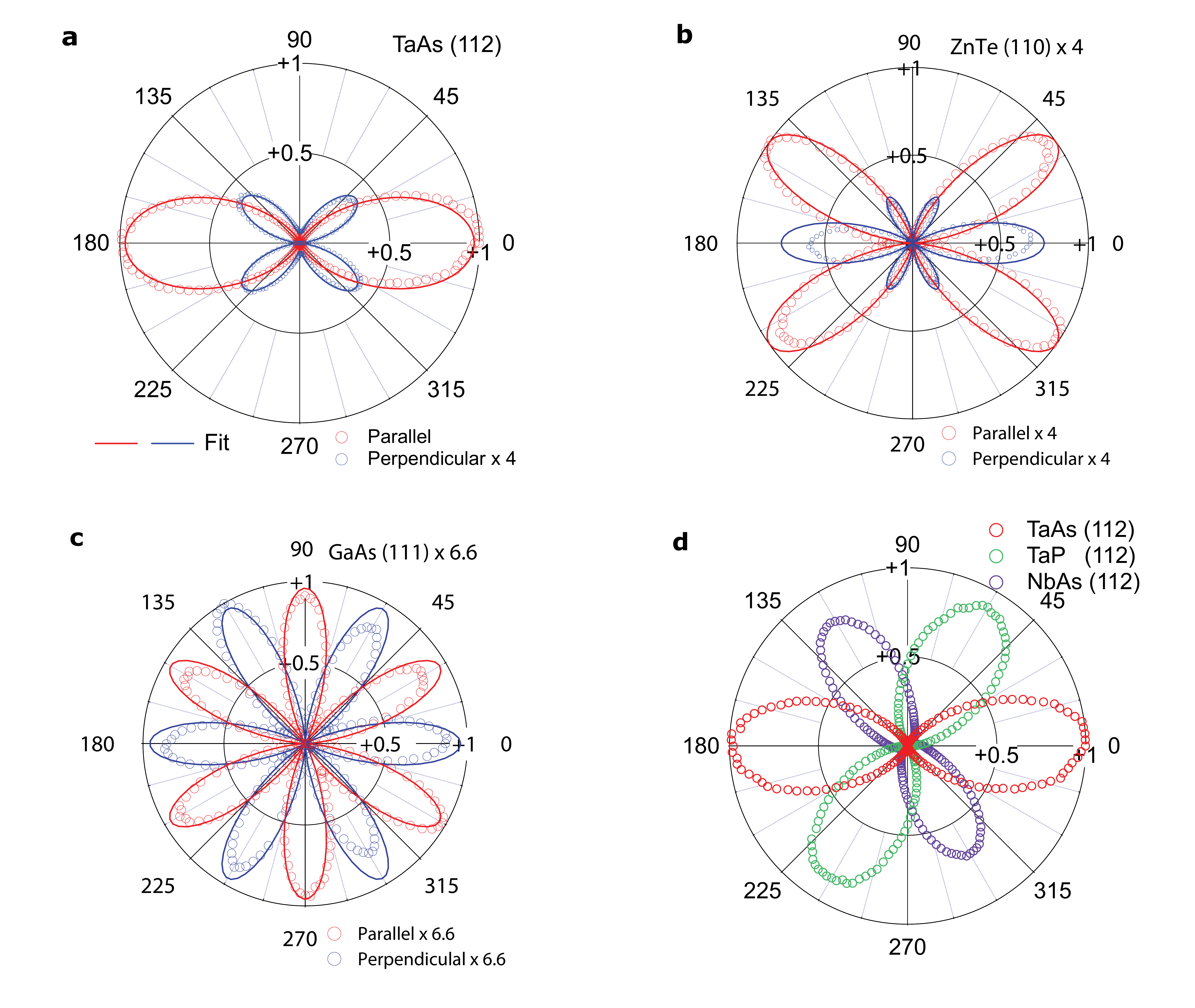}
\caption{   \textsf{\textbf{Benchmark SHG experiments on TaAs (112), TaP (112), NbAs (112), ZnTe (110) and GaAs (111).}}  SHG polar plots in the same scale for TaAs (112) (\textsf{\textbf{a}}), ZnTe (110) (\textsf{\textbf{b}}), and GaAs (111) (\textsf{\textbf{c}}) in both the parallel and perpendicular generator/analyzer configurations at room temperature. For TaAs data in the perpendicular configuration is magnified by a factor of 4 for clarity. The SHG intensity of ZnTe and GaAs are multiplied by a factor 4 and 6.6 respectively to match the peak value of TaAs. (\textsf{\textbf{d}}), SHG polar plots for TaAs (112), TaP (112) and NbAs (112) in the parallel configurations at room temperature with  plots of TaP and NbAs rotated by 60$^\circ$ and 120$^\circ$ for clarity. }
 \label{Fig3}
\end{figure*}

Figs. \ref{Fig3}a-c show polar plots of SHG intensity as TaAs (112), ZnTe (110), and GaAs (111) are (effectively) rotated about the optic axis with the generator and analyzer set at 0$^\circ$ and 90$^\circ$.  Also shown (as solid lines) are fits to the polar patterns obtained by rotating the $\chi^{(2)}$ tensor to a set of axes that includes the surface normal, which is (110) and (111) for our ZnTe and GaAs crystals respectively (Methods and Supplementary Information section A). Even prior to analysis to extract the ratio of $d$ coefficients between the various crystals, it is clear that the SHG response of TaAs (112) is large, as the peak intensity in this geometry exceeds ZnTe (110) by a factor of 4.0($\pm0.1$) and GaAs (111) by a factor of 6.6($\pm0.1$). Fig. \ref{Fig3}d compares the parallel polarization data for TaAs shown in Fig. \ref{Fig3}a with SHG measured under the same conditions in other TMMP's TaP and NbAs (112) facets. The strength of SHG from the three crystals, which share the same 4$mm$ point group, is clearly very similar, with TaP and NbAs intensities relative to TaAs of 0.90($\pm$0.02) and 0.76($\pm$0.04), respectively. The SHG response in these compounds is also dominated by the $d_{33}$ coefficient. Finally, we found that the SHG intensity of all three compounds does not decrease after exposure to atmosphere for several months.

\begin{table}[h!]
\begin{tabular}{ |p{1.2cm}||p{0.8cm}|p{1.8cm}|p{2.4cm}|p{1.7cm}|}
 \hline
 \hline
  Material  & $|d_{ij}|$ &  $|d|$ ({pm/V}) & Fundamental wavelength (nm) & Reference  \\
 \hline
 TaAs & $d_{33}$  & 3600 ($\pm 550$) & 800 & This work \\
GaAs & $d_{14}$ & 350$^*$ & 810 & Ref. \onlinecite{bergfeld2003second} \\
ZnTe & $d_{14}$ & 250, 450$^*$ & 800, 700 & Ref. \onlinecite{wagner1998dispersion} \\
BaTiO$_3$ & $d_{33}$  & 15 & 900 & Ref. \onlinecite{miller1964optical} \\
BiFeO$_3$ & $d_{33}$  & 15-19 & 1550, 800 & Ref. \onlinecite{haislmaier2013large, ju2009electronic}$^t$ \\
LiNbO$_3$ & $d_{33}$  & 26 & 852 & Ref. \onlinecite{shoji1997absolute} \\
BiFeO$_3$ & $d_{33}$  & 130$^*$ & 500 & Ref. \onlinecite{ju2009electronic}$^t$\\
BaTiO$_3$ & $d_{33}$  & 100$^*$ & 170 & Ref. \onlinecite{young2012first}$^t$\\
PbTiO$_3$ & $d_{33}$  & 200$^*$ & 150 & Ref. \onlinecite{young2012first}$^t$\\

 \hline
\end{tabular}
\caption{Second harmonics generation coefficients of different materials at room temperature. Second harmonic optical susceptibility can be calculated by $\chi_{ijk}=2d_{ij}$. $^*$ denotes the peak value of the material. $^t$ denotes theoretical calculation. The uncertainty of $d_{33}$ in TaAs is determined by setting $d_{15}$ and $d_{31}$ in and out of phase to $d_{33}$ in the fit.}
\label{tab2}
\end{table}

To obtain the response of TMMP's relative to the two benchmark materials we used the Bloembergen-Pershan formula \cite{bloembergen1962light1} to correct for the variation in specular reflection of SH light that results from the small differences in the index of refraction of the three materials at the fundamental and SH frequency. (See Methods. Details concerning this correction, which is less than 20$\%$, can be found in Supplementary Information section B). Table \ref{tab2} presents the results of this analysis, showing that $|d_{33}|\cong3600$ pm/V at fundamental wavelength 800 nm in TaAs exceeds values in benchmark materials GaAs \cite{bergfeld2003second} and ZnTe \cite{wagner1998dispersion} by approximately one order of magnitude, even when measured at wavelengths where their response is largest. The $d$ coefficient in TaAs at 800 nm exceeds corresponding values in ferroelectric materials BiFeO$_3$ \cite{haislmaier2013large}, BaTiO$_3$ \cite{miller1964optical}, LiNbO$_3$ \cite{shoji1997absolute} by two orders of magnitude. In the case of the ferroelectric materials, SHG measurements have not been performed in their spectral regions of strong absorption, typically 3-7 eV. However, \textit{ab-initio} calculations consistently predict that the resonance enhanced $d$ values in this region do not exceed roughly 500 pm/V\cite{ju2009electronic, young2012first}.

The results described above raise the question of why $\chi_{zzz}$ in the TMMP's is so large. Answering this question quantitatively will require further work in which measurements of $\chi^{(2)}$ as a function of frequency are compared with theory based on \textit{ab initio} bandstructure and wavefunctions. For the present, we describe a calculation of $\chi^{(2)}$ using a minimal model of a WSM that is based on the approach to nonlinear optics proposed by Morimoto and Nagaosa (MN) \cite{Morimoto2015}. This theory clarifies the connection between bandstructure topology and SHG, and provides a concise expression with clear geometrical meaning for $\chi^{(2)}$. Hopefully this calculation will motivate the \textit{ab initio} theory that is needed to quantitatively account for the large SH response of the TMMP's and its possible relation to the existence of Weyl nodes.

The MN result for the dominant ($zzz$) response function is,

\begin{align}
\t{Re}\{\sigma^{(2)}_{zzz}(\omega,2\omega) \}
&\cong
\frac{\pi e^3}{2\hbar \omega^2}
\int \frac{d^3 \bm k}{(2\pi)^3}
|v_{z,12}|^2
R_{zz}(\bm k)
\n
&\quad \times
\left[
-\delta(\epsilon_{21}-\hbar\omega)
+\frac{1}{2}\delta(\epsilon_{21}-2\hbar\omega)
\right].
\label{Eq1}
\end{align}
In Eq. 1 the nonlinear response is expressed as a second-order conductivity, $\sigma_{zzz}(\omega,2\omega)$, relating the current induced at $2\omega$ to the square of the applied electric field at $\omega$, \textit{i.e}., $J_z(2\omega)=\sigma_{zzz} E_z^2(\omega)$. (The SH susceptibility is related to the conductivity through the relation $\chi^{(2)}=\sigma^{(2)}/2i\omega\epsilon_0$). The indices 1 and 2 refer to the valence and conduction bands, respectively, $\epsilon_{21}$ is the transition energy, and $v_{i,12}$ is the matrix element of the velocity operator $v_i=(1/\hbar)\partial H/\partial k_i$. Bandstructure topology appears in the form of the ``shift vector," $R_{zz}\equiv\partial_{k_z} \varphi_{z,12}+a_{z,1}-a_{z,2}$, which is a gauge-invariant length formed from the $k$ derivative of the phase of the velocity matrix element, $\varphi_{12}=\t{Im}\{\log v_{12}\}$, and the difference in Berry connection, $a_i=-i\bra{u_n}\partial_{k_i} \ket{u_n}$, between bands 1 and 2. Physically, the shift vector is the $k$-resolved shift of the intracell wave function for the two bands
connected by the optical transition.

We consider the following minimal model for a time-reversal symmetric WSM that supports four Weyl nodes,

\begin{align}
H&=
t \Big\{
[\cos k_x a
+m_y(1-\cos k_y a)
+m_z(1-\cos k_z a)
]\sigma_x
\n
&\qquad
+ [\sin k_y a +\Delta \cos (k_y a) s_x]\sigma_y
+\sin (k_z a) s_x\sigma_z
\Big\}.
\label{Eq2}
\end{align}
Here, $\sigma_i$ and $s_i$ are Pauli matrices acting on orbital and spin degrees of freedom, respectively, $t$ is a measure of the bandwidth, $a$ is the lattice constant, $m_y$ and $m_z$ are parameters that introduce anisotropy, and inversion breaking is introduced by $\Delta$. The Hamiltonian defined in Eq. \ref{Eq2} preserves two-fold rotation symmetry about the $z$-axis and the mirror symmetries $M_x$ and $M_y$. These symmetries form a subset of $4mm$ point group which is relevant to optical properties of TMMP's.

\begin{figure*}[htp]
\includegraphics[width=\textwidth]{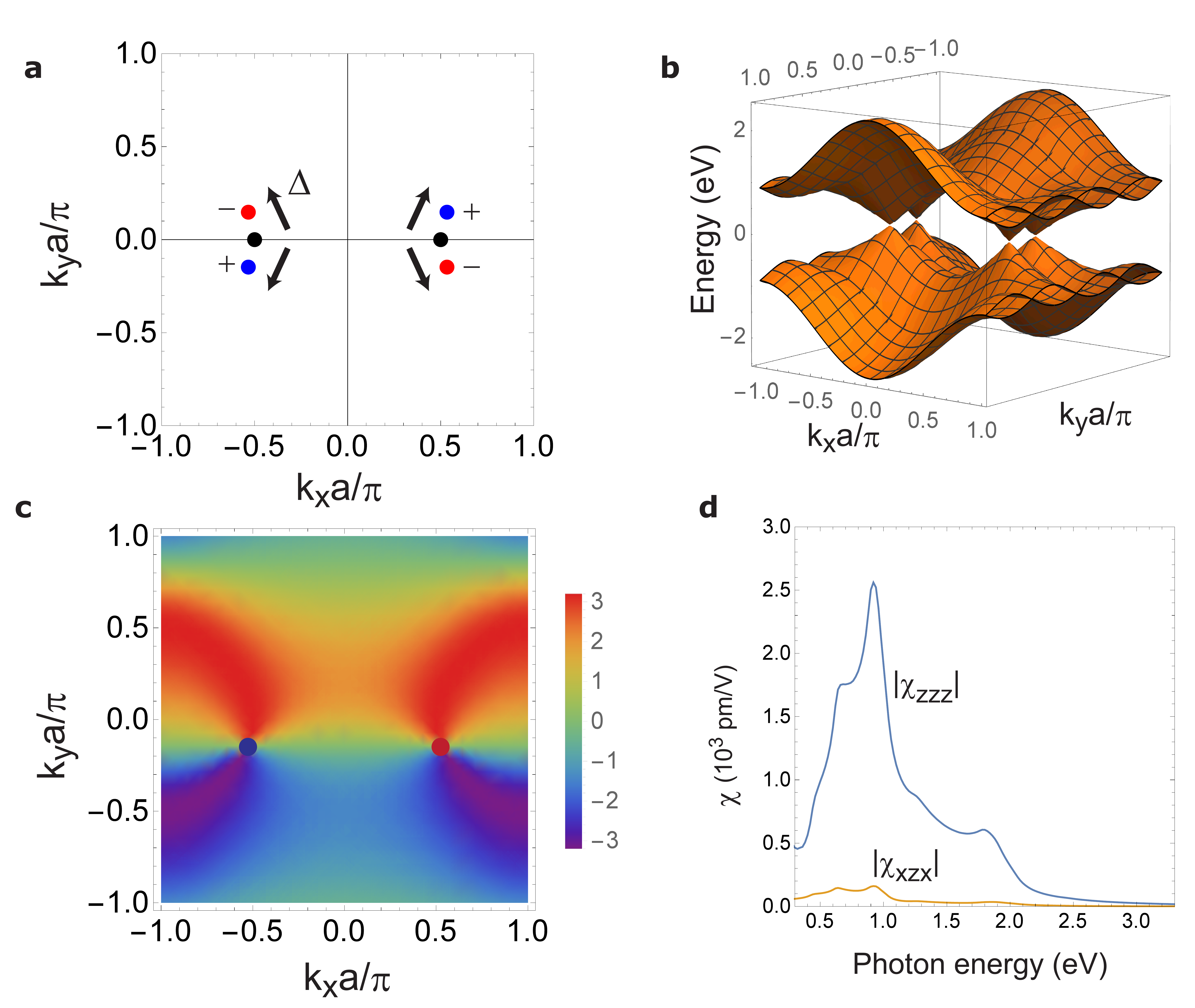}

\caption{  \textsf{\textbf{Numerical results for the second harmonic response of a Weyl semimetal.}}
(\textsf{\textbf{a}}) Location of Weyl points in the $k_z=0$ plane for $\Delta=0.5$.
For $\Delta=0$, Weyl points with opposite chiralities are located at
$(\pm\pi/2a,0,0)$ (denoted by black dots).
(\textsf{\textbf{b}}) The band structure for $\Delta=0.5$ and $k_z=0$.
(\textsf{\textbf{c}}) Color plot of $|v_{12}|^2 R_{zz}$ which appears as an integrand in the formula for $\sigma^{(2)}_{zzz}$. For clarity we plot only $|v_{12}|^2 R_{zz}$ for $s_x=+1$ bands with Weyl points located at $k_y<0$.
$|v_{12}|^2 R_{zz}$ for $s_x=-1$ bands is obtained by setting $\bm k \to -\bm k$.
$|v_{12}|^2 R_{zz}$ shows structures at Weyl points.
(\textsf{\textbf{d}}) $|\chi_{zzz}|$ and $|\chi_{xzx}|$ plotted as a function of incident photon energy for $\Delta=0.5$. We adopted parameters $t=0.8 \t{ eV}, m_y=1, m_z=5$.
}
 \label{Fig4}
\end{figure*}

Fig. \ref{Fig4} illustrates the energy levels, topological structure, and SHG spectra that emerge from this model. As shown in Fig. \ref{Fig4}a, pairs of Weyl nodes with opposite chirality overlap at two points, $\bm k=(\pm \pi/2a, 0, 0)$, in the inversion-symmetric case with $\Delta=0$. With increasing $\Delta$ the nodes displace in opposite directions along the $k_y$ axis, with $\Delta k_y \cong \Delta/a$. The energy of electronic states in $k_z=0$ plane, illustrating the linear dispersion near the four Weyl points, is shown in Fig. \ref{Fig4}b. Fig. \ref{Fig4}c shows the corresponding variation of $|v_{12}|^2 R_{zz}(\bm k)$ for the $s_x=+1$ bands whose Weyl points are located at $k_y < 0$ (the variation of $|v_{12}|^2 R_{zz}(\bm k)$ for the $s_x=-1$ bands is obtained from the transformation $\bm k \rightarrow -\bm k$). The magnitude of $\sigma^{(2)}$ derived from this model vanishes as $\Delta\rightarrow 0$ and is also sensitive to the anisotropy parameters $m_y$ and $m_z$. Fig. \ref{Fig4}d shows that spectra corresponding to parameters $t=0.8$ eV, $\Delta=0.5$, $m_z=5$, and $m_y=1$ can qualitatively reproduce the observed amplitude and large anisotropy of $\chi^{(2)}(\omega,2\omega)$.

As discussed above, our minimal model of an inversion-breaking WSM is intended mainly to motivate further research into the mechanism for enhanced SHG in the TMMP's. However, the model does suggest universal properties of $\chi^{(2)}$ that arise from transitions near Weyl nodes between bands with nearly linear dispersion. According to bulk bandstructure measurements \cite{YangNatPhys2015}, such transitions are expected at energies below approximately 100 meV in TaAs family, corresponding to far-infrared and THz regimes. In this regime, where the interband excitation is within the Weyl cones, the momentum-averaged $|v_{12}|^2 R_{zz}(\bm k)$ tends to nonzero value, $\langle v^2 R\rangle$, leading to the prediction that $\sigma^{(2)}\rightarrow g(\omega) \langle v^2 R\rangle/\omega^2$ as $\omega\rightarrow 0$. Because $g(\omega)$, the joint density of states for Weyl fermions is proportional to $\omega^2$, we predict that $\sigma^{(2)}$ approaches a constant (or alternatively $\chi^{(2)}$ diverges as $1/\omega$) as $\omega\rightarrow 0$, even as the linear optical conductivity vanishes in proportion to $\omega$ \cite{hosur2012charge}. The $1/\omega$ scaling of SHG and optical rectification is a unique signature of a WSM in low-energy electrodynamics, as it requires the existence of both inversion breaking and point nodes. In real materials, this divergence will be cutoff by disorder and nonzero Fermi energy. Disorder induced broadening, estimated from transport scattering rates  \cite{xu2016optical1} and Pauli blocking from nonzero Fermi energy, estimated from optical conductivity \cite{xu2016optical1} and band calculation \cite{WengPRX2015}, each suggests a low energy cutoff in the range of a few meV.

We conclude by observing that the search for inversion-breaking WSMs has led, fortuitously, to a new class of polar metals with unusually large second-order optical susceptibility.  Although WSM€'s are not optimal for frequency doubling applications in the visible regime because of their strong absorption, they are promising materials for THz generation and opto-electronic devices such as far-infrared detectors because of their unique scaling in the $\omega\rightarrow 0$ limit. Looking forward, we hope that our findings will stimulate further investigation of nonlinear optical spectra in inversion-breaking WSM's for technological applications and in order to identify the defining response functions of Weyl fermions in crystals.

\section{Methods}

\textbf{Crystal growth and structure characterization}

Single crystals of TaAs, TaP and NbAs were grown by vapor transport with iodine as the transport agent. First, polycrystalline TaAs/TaP/NbAs was produced by mixing stoichiometric amounts of Ta/Nb and As/P and heating the mixture to 1100/800/700 $^{\circ}$C in an evacuated quartz ampule for 2 days. 500 mg of the resulting powder was then resealed in a quartz ampoule with 100 mg of iodine and loaded into a horizontal two-zone furnace. The temperature of the hot and cold ends were held at 1000 $^{\circ}$C and 850 $^{\circ}$C, respectively, for TaAs and 950 $^{\circ}$C and 850 $^{\circ}$C for TaP and NbAs. After 4 days well-faceted crystals up to several millimeters in size were obtained. Crystal structure was confirmed using single-crystal x-ray mico-Laue diffraction at room temperature at beamline 12.3.2 at the Advanced Light Source.

\textbf{Optics setup for second harmonic generation}
The optical set up for measuring SHG is illustrated in Fig. \ref{Fig1}a. Generator pulses of 100 fs duration  and center wavelength 800 nm pass through a mechanical chopper that provides amplitude modulation at 1 kHz and are focused at near normal incidence onto the sample. Polarizers and waveplates in the beam path are used to vary the direction of linear polarization and to generate circularly polarization. Both the specularly reflected fundamental and the second harmonic beam are collected by a pickoff mirror and directed to a short-pass, band-pass filter combination that allows only the second harmonic light to reach photomultiplier tube (PMT) photodetector. Another wire-grid polarizer placed before the PMT allows for analysis of the polarization of the second harmonic beam. Temperature dependence measurement was performed by mounting TaAs sample in a cold-finger cryostat on a xyz-micrometer stage. Benchmark measurements on TaAs, TaP, NbAs, ZnTe and GaAs were performed at room temperature in atmosphere with the samples mounted on a xyz-micrometer stage to maximize the signal.

\textbf{Calculation and fitting procedure for SHG}
We first fit data obtained with fixed analyzer at 90$^{\circ}$, which is the [1,-1,0] crystal axis, because there is only one free parameter ($|d_{15}|$) in this configuration. With this value for $|d_{15}|$, we then fit data in the three other types of scans discussed in the text with $d_{15}$ and $d_{31}$ set in and out-of phase with $d_{33}$. This procedure yields upper and lower bounds on the anisotropy ratios $|d_{15}/d_{33}|$ and $|d_{31}/d_{33}|$ that are shown with error bars in Fig. \ref{Fig2}(d). In the case of GaAs and ZnTe all scans are fit accurately by the only symmetry allowed free parameter, which is $|d_{14}|$. See Supplementary Information for more details.

\textbf{Bloembergen-Pershan correction}
When measuring in the reflection geometry, one need to consider the boundary condition to calculate $\chi^{(2)}$ from $\chi^{(2)}_R$ that was directly measured, where `R' stands for reflection geometry. The correction was worked out by Bloembergen-Pershan (BP) (See supplementary information):

\begin{equation}
 \begin{split}
 \chi^{(2)}_{R} \equiv&-\frac{E_{R}(2\omega)}{\epsilon_0 E(\omega)^2}  \\
  =&\frac{\chi^{(2)}}{(\epsilon^{1/2}(2\omega)+\epsilon^{1/2}(\omega))(\epsilon^{1/2}(2\omega)+1)}T(\omega)^2.
  \end{split}
\end{equation}

\normalsize

\noindent where $\epsilon$ is the relative dielectric constant and $T(\omega)=\frac{2}{ n(\omega)+1}$ is the Fresnel coefficient of the fundamental light. In the current experiment performed at 800 nm, the BP correction is quite small (less than 20 \%). See Supplementary Information for more details.


\section{addendum}
We thank B. M. Fregoso, T. R. Gordillo, J. Neaton and Y. R. Shen for helpful discussions and B. Xu for sharing refractive index data of TaAs. Measurements and modeling were performed at the Lawrence Berkeley National Laboratory in the Quantum Materials program supported by the Director, Office of Science, Office of Basic Energy Sciences, Materials Sciences and Engineering Division, of the U.S. Department of Energy under Contract No. DE-AC02-05CH11231. J.O., L.W., and A.L. received support for performing and analyzing optical measurements from the Gordon and Betty Moore Foundation's EPiQS Initiative through Grant GBMF4537 to J.O. at UC Berkeley.  Sample growth was supported by the Gordon and Betty Moore Foundation's EPiQS Initiative Grant GBMF4374 to J.A. at UC Berkeley. T.M. is supported by the Gordon and Betty Moore Foundation's EPiQS Initiative Theory Center Grant GBMF4307 to UC Berkeley. J.E.M. received support for travel from the Simons Foundation. The authors would like to thank Nobumichi Tamura for his help in performing crystal diffraction and orientation on beamline 12.3.2 at the Advanced Light Source. N. Tamura and the ALS are supported by the Director, Office of Science, Office of Basic Energy Sciences, of the U.S. Department of Energy under Contract No. DE-AC02-05CH11231. J. A. and N. N. acknowledge support by the Office of Naval Research under the Electrical Sensors and Network Research Division, Award No. N00014-15-1-2674.

\textit{Competing Interests: }The authors declare that they have no
competing financial interests.

\textit{Correspondence: }Correspondence and requests for materials
should be addressed to L.W. (email:liangwu@berkeley.edu) and J.O. (email:jworenstein@lbl.gov).

\section{Author Contribution}
L.W. and J.O. conceived the project. L.W. and S.P. performed  and contributed equally to the SHG measurements with assistance from E. T. and A. L..  L.W. and J.O. analyzed the data. T.M. and J.M. performed the model calculation. L.W., T.M. and J.O. performed the frequency scaling analysis.  N.N. and J.A. grew the crystals and characterized the crystal structure.  L.W., T.M. and J.O. wrote the manuscript. All authors commented on the manuscript.


\begin{widetext}

\newpage

\clearpage
\newpage
\bigskip

\newpage
\bigskip
\setcounter{figure}{0}
\setcounter{equation}{0}
\section{S\lowercase{upplementary} I\lowercase{nformation for} ``G\lowercase{iant anisotropic nonlinear optical response in transition metal monopnictide \uppercase{W}eyl semimetals}''}

\subsection{Dependence of SHG amplitude on polarization for different crystal and surfaces used in this study}
\label{SIsection1}
In general, the polarization, \textbf{P}, in materials has contributions from higher orders of the electric field, \textbf{E}, in addition to the linear response, such that,

\begin{equation}
\mathbf{P}=\mathbf{P_0}+\epsilon_0\chi_e \mathbf{E}+ \epsilon_0\chi^{(2)}\mathbf{E}^2+\cdot\cdot\cdot
\end{equation}

\noindent In noncentrosymmetric materials the second order term, $P_i^{(2)}=\epsilon_{0}\chi_{ijk}E_j E_k$, that gives rise to frequency mixing, SHG, and optical rectification, is allowed\cite{Boyd2, shen1984principles}. The latter two phenomena arise from excitation with a single frequency, such that there is an automatic symmetry with respect to permutation of the second and third indices. This motivates the use of a $3\times 6$ second rank tenor $d_{ij}$ instead of $\chi_{ijk}$\cite{Boyd2}. The relation between  $d_{ij}$ and $\chi_{ijk}$ is as follows: the first index $i=1, 2, 3$ in $d_{ij}$ corresponds to $i'=x, y, z$ respectively in $\chi_{i'j'k'}$ and the second index $j=1, 2, 3, 4, 5, 6$ in $d_{ij}$ corresponds to $j'k'=xx, yy, zz, yz/zy, zx/xz, xy/yx$ in $\chi_{i'j'k'}$\cite{Boyd2}. In terms of the $d$ tensor, the relation between second order polarization and electric field has the form\cite{Boyd2}:
\begin{equation}
\begin{bmatrix}
    P_{1}(2\omega)  \\
   P_{2}(2\omega)  \\
   P_{3}(2\omega)
  \end{bmatrix}
=
  2\epsilon_{0}\begin{bmatrix}
    d_{11} & d_{12} & d_{13} & d_{14} & d_{15} & d_{16} \\
    d_{21} & d_{22} & d_{23} & d_{24} & d_{25} & d_{26} \\
    d_{31} & d_{32} & d_{33} & d_{34} & d_{35} & d_{36} \\
  \end{bmatrix}
  \begin{bmatrix}
    E_{1}^2(\omega)  \\
   E_{2}^2(\omega)  \\
   E_{3}^3(\omega) \\
    2E_{2}(\omega)E_{3}(\omega)  \\
   2E_{1}(\omega)E_{3}(\omega)  \\
   2E_{1}(\omega)E_{2}(\omega)
  \end{bmatrix}.
  \label{Eq_S2}
\end{equation}

For a crystal with effective point group symmetry $4mm$, the nonzero elements are $d_{15}=d_{24}$,  $d_{31}=d_{32}$ and $d_{33}$. Note that transition metal monopnictides (TMMPs) such as TaAs belong to the non-symmorphic $I4_1md$ space group which has screw rotation instead of C$_4$ rotation. However, screw and C$_4$ rotation symmetries lead to the same constraints on $\chi^{(2)}$ in optics and therefore SHG is described by the $4mm$ point group in TMMPs. Predictions based on Eq. \ref{Eq_S2} for the angular dependence of the SHG intensity for four scans that involve linear polarized light normally incident on the TaAs or TaP or NbAs (112) surface are given below.  Eqs. \ref{Eq_S3} and \ref{Eq_S4} refer to scans where two polarizers are synchronously rotated to simulate rotation of the sample. Note that we omitted the constant of 2$\epsilon_{0}$ in the following calculations. $I_{para}$ and $I_{perp}$ correspond to generator and analyzer polarization set parallel and perpendicular, respectively,

\begin{equation}
I_{para}(\theta_1)=\frac{1}{27}|(d_{33}+4d_{15}+2d_{31})\cos^3\theta_1+3(2d_{15}+d_{31})\sin^2\theta_1\cos\theta_1|^2,
\label{Eq_S3}
\end{equation}

\begin{equation}
I_{perp}(\theta_1)=\frac{1}{27}|(d_{33}-2d_{15}+2d_{31})\cos^2\theta_1\sin\theta_1+3d_{31}\sin^3\theta_1|^2.
\label{Eq_S4}
\end{equation}

\noindent Eqs. \ref{Eq_S5} and \ref{Eq_S6} refer to scans where the analyzer is fixed at $0^{\circ}$ (parallel to [1,1-1] crystal axis) and $90^{\circ}$ (parallel to [1,-1,0] crystal axis), respectively, and the direction of linear polarization of the generator is scanned,

\begin{equation}
I_{0^{\circ}}(\theta_1)=\frac{1}{27}|(d_{33}+4d_{15}+2d_{31})\cos^2\theta_1+3d_{31}\sin^2\theta_1|^2,
\label{Eq_S5}
\end{equation}

\begin{equation}
I_{90^{\circ}}(\theta_1)=\frac{1}{3}|d_{15}|^2\sin^{2}(2\theta_1).
\label{Eq_S6}
\end{equation}

The remaining scan that provides independent information constraining the elements of the $d$ tensor is a measurement of SHG circular dichroism, which is the difference in SHG intensity with incident light of left and right circular polarization. To lowest order in $d_{ij}$,

\begin{equation}
I_{\sigma^{+}}(\theta_2)-I_{\sigma^{-}}(\theta_2)=\frac{2}{9}\textrm{Im}\{d_{15}d_{33}^{*}\}\sin2\theta_2.
\end{equation}
The circular dichroism in SHG is proportional to $Im{(d_{33}d_{15}^\ast)}$ and therefore measures the component of $d_{15}$ that is out-of-phase with $d_{33}$. From comparison of the circular dichroism and linear polarization measurements we find a relative phase of 30 $^\circ$ ($\pm$ 10 $^\circ$) between $d_{15}$ and $d_{33}$ at 300 K.

The benchmark materials, GaAs and ZnTe, have the same point group \textit{$\bar{4}$3m}, and the only nonzero components of the $d$ tensor are $d_{14}=d_{25}=d_{36}$. For light normally incident on the ZnTe (110) surface, the SHG angular dependencies are:

\begin{equation}
I_{para}(\theta_1)=9|d_{14}|^2\cos^4\theta_1\sin^2\theta_1,
\end{equation}

\begin{equation}
I_{perp}(\theta_1)=|d_{14}|^2(2\cos\theta_1\sin^2\theta_1-\cos^3\theta_1)^2.
\end{equation}

\noindent For GaAs (111) they are:

\begin{equation}
I_{para}(\theta_1)=\frac{2}{3}|d_{14}|^2(cos^3\theta_1-3\cos\theta_1\sin^2\theta_1)^2.
\end{equation}

\begin{equation}
I_{perp}(\theta_1)=\frac{2}{3}|d_{14}|^2(\sin^3\theta_1-3\cos^2\theta_1\sin\theta_1)^2.
\end{equation}

\subsection{Obtaining nonlinear response coefficients from benchmark materials}
\label{SIsection2}
From Eq. 3, the peak SHG intensity from TaAs (112) is proportional to $|d_{\textrm{eff}}|^2/27$, where $d_{\textrm{eff}}\equiv d_{33}+4d_{15}+2d_{31}$. Peak SHG intensities from ZnTe (110) and GaAs (111) are proportional to $4|d_{14}|^2/3$ and $2|d_{14}|^2/3$, respectively. When measured in reflection, in addition to the nonlinear response coefficients, the SHG intensity depends on the index of refraction at the fundamental and the second harmonic frequencies. To compare SHG intensities from different compounds we use a formula derived by Bloembergen and Pershan \cite{bloembergen1962light},

\small
\begin{equation}
\chi^{(2)}_{R} \equiv-\frac{E_{R}(2\omega)}{\epsilon_0 E(\omega)^2}=\frac{\chi^{(2)}}{(\epsilon^{1/2}(2\omega)+\epsilon^{1/2}(\omega))(\epsilon^{1/2}(2\omega)+1)}T(\omega)^2.
  \label{Eq1}
\end{equation}
\normalsize

\noindent where $\epsilon$ is the relative dielectric constant, `R' stands for reflection geometry and $T(\omega)=\frac{2}{ n(\omega)+1}$ is the Fresnel coefficient of the fundamental light. Values for index of refraction were taken from the literature: ZnTe\cite{li1984refractive}, GaAs\cite{forouhi1988optical} and complex index of TaAs was measured by the method that can be found in Ref. \onlinecite{xu2016optical}. The correction factor turns out to be small, less than 20\%, because the indices of refraction for the three compounds are similar in magnitude (within 30 $\%$ difference) at the relevant frequencies. Without taking account of this correction, we obtain $|d_{33}| \sim 3000 \pm450$ pV/m for TaAs by referring to the value of ZnTe\cite{wagner1998dispersion2}.

After establishing the relative size of the $d$ coefficients by the procedure described above, we used measurements of the absolute amplitude of $d_{14}$ in ZnTe \textit{et al.}\cite{wagner1998dispersion2} to obtain absolute amplitudes for GaAs and TaAs.  As reported in the main text, we obtain $|d_{33}|\sim$ 3600 pV/m in TaAs. Using the same procedure of referencing to ZnTe, we obtain $|d_{14}| \sim 380$ pV/m for GaAs, which agrees with the literature value\cite{bergfeld2003second2}.

\end{widetext}

\end{document}